# Does Basis Set Superposition Error Significantly Affect Post-CCSD($T$) Corrections?

Vladimir Fishman | Emmanouil Semidalas | Margarita Shepelenko | Jan M. L. Martin

Department of Molecular Chemistry and Materials Science, Weizmann Institute of Science, Rehovot, Israel

**Correspondence:** Jan M. L. Martin (gershom@weizmann.ac.il)

Received: 24 September 2024 | Revised: 29 November 2024 | Accepted: 29 November 2024

**Funding:** This work was supported by Israel Science Foundation (1969/20).

**ABSTRACT**

We have investigated the title question for both a subset of the W4-11 total atomization energies benchmark, and for the A24x8 noncovalent interactions benchmark. Overall, counterpoise corrections to post-CCSD($T$) contributions are about two orders of magnitude less important than those to the CCSD($T$) interaction energy. Counterpoise corrections for connected quadruple substitutions ($Q$) are negligible, and $(Q)_\Lambda - (Q)$ or $T_4 - (Q)$ especially so. In contrast, for atomization energies, the $T_3 - (T)$ counterpoise correction can reach about 0.05 kcal/mol for small basis sets like cc-pVDZ, thought it rapidly tapers off with cc-pVTZ and especially aug-cc-pVTZ basis sets. It is reduced to insignificance by the extrapolation of $T_3 - (T)$ applied in both W4 and HEAT thermochemistry protocols. In noncovalent dimers, the differential BSSE on post-CCSD($T$) correlation contributions is negligible even in basis sets as small as the unpolarized split-valence cc-pVDZ(no d).

## 1 | Introduction

Basis set superposition error (BSSE) is the error in the interaction energy between, for example, a dimer $AB$ and its constituent monomers $A$ and $B$ when evaluated in a finite basis set. (At the complete basis set (CBS) limit, BSSE vanishes.) The classic remedy for BSSE is the Boys–Bernardi counterpoise (CP) method [1].

$$\begin{aligned} \mathrm{BSSE} &= E[A] + E[B] - E[A(B)] - E[B(A)] \\ &= E[AB] - E[A(B)] - E[B(A)] - (E[AB] - E[A] - E[B]) \\ &= D_e[\mathrm{raw}] - D_e[\mathrm{CP}] \end{aligned} \quad (1)$$

where $D_e$ denotes the dissociation energy, $E[AB]$ is the total energy of the dimer, $E[A(B)]$ the total energy of $A$ in the presence of the basis functions on $B$, $E[A]$ the corresponding total energy in their *absence*, and so forth.

Inclusion of BSSE in noncovalent interaction (NCI) studies is more or less standard operating procedure, especially in smaller and medium basis sets, as the CP corrections may be on the same order or magnitude as the interaction energies of interest.

Now it is indeed true that full counterpoise does *not* guarantee hewing closer to the CBS limit: as shown by Burns, Marshall, and Sherrill [3] for orbital WFT calculations, and by Brauer, Kesharwani, and Martin [2] for explicitly correlated [4–6] F12 calculations, error compensation may take place between BSSE (which always overbinds) and IBSI (intrinsic basis set incompleteness, which almost invariably underbinds). Hence, for small basis sets, complete neglect of BSSE may actually be beneficial, and for medium-size basis sets, "half-counterpoise" (average of corrected and uncorrected interaction energies) tends to offer superior performance [2, 3, 7]. See Figure 1 for an illustration.

In computational thermochemistry, however, the IBSI overwhelms BSSE to such an extent that most researchers make no effort to apply BSSE corrections. This is particularly the case for total atomization energies (TAE), which are the quantum chemical "cognates" of heats of formation $\Delta H_f^\circ$. The "raw" and

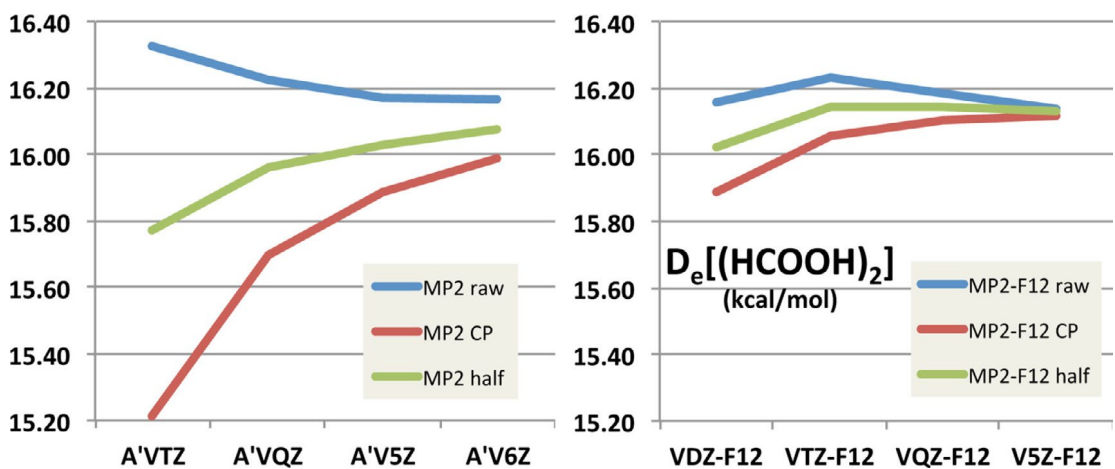

**FIGURE 1** | Illustration of effect of BSSE on basis set convergence for formic acid dimer. Reprinted from figure 1 in Reference [2], with Creative Commons license.

CP-corrected TAEs are defined analogously to the NCI situation as (e.g., for a triatomic):

$$TAE_{\text{raw}}[ABC] = E[A] + E[B] + E[C] - E[ABC]$$
$$TAE_{\text{CP}}[ABC] = E[A(BC)] + E[(A)B(C)] + E[(AB)C] - E[ABC]$$
$$\Delta_{\text{CP}} TAE_{\text{CP}}[ABC] = TAE_{\text{raw}}[ABC] - TAE_{\text{CP}}[ABC] \quad (2)$$
$$= E[A] + E[B] + E[C] - E[A(BC)] - E[(A)B(C)] - E[(AB)C]$$

where the Wells and Wilson [8] SSFC (site-site function counterpoise) $n$-body generalization of Equation (1) has been applied. (This effectively amounts to evaluating each atomic energy in the full molecular basis set.)

Higher-accuracy computational thermochemistry protocols like ccCA [9, 10] by the Wilson group, HEAT by the Stanton group [11–14], Weizmann-$n$ by our own group [15–17], and FPD (Feller–Peterson–Dixon, see References [18, 19] and references therein) all entail some variant of basis set extrapolation. If the latter works properly, it ought to eliminate BSSE altogether. (Indeed, we recently [20] exploited this fact to "reverse-engineer" basis set extrapolations.)

Studies of noncovalent interactions, with rare exceptions (such as References [21–23]) stick to the CCSD($T$) [24, 25] "gold standard of quantum chemistry" and ignore post-CCSD($T$) corrections. However, in thermochemistry, especially for TAEs, it is well-known (see References [11, 26] for early reports) that kJ/mol accuracy cannot be achieved without them. CCSD($T$) in fact outperforms the more rigorous CCSDT owing to a well-established error compensation (e.g., References [11, 13, 15, 26]): higher-order triples, $T_3 - (T)$, are almost always antibonding, while connected quadruples ($Q$) are universally bonding.

On the one hand, CCSDT($Q$) and especially CCSDTQ have very steep CPU time scalings of $O(n_{\text{occ}}^4 N_{\text{virt}}^5)$ and $O(n_{\text{occ}}^4 N_{\text{virt}}^6)$, respectively. On the other hand, these higher-order corrections converge much more rapidly with the basis set than the overall correlation energy [16]. In response to a reviewer comment, we offer Table 1 as an illustration, compiled from data in the supporting information of Reference [28]. The statistics given there cover a 65-molecule subset of the 200-molecule W4-17 thermochemical benchmark [27]; the subset spans a broad range of nondynamical (static) correlation character, from essentially pure dynamical correlation in $H_2O$ and $CH_4$ at one end, to strong static correlation in $O_3$, singlet $C_2$, and BN at the other end. It is clearly seen in Table 1 that RMS contributions taper off rapidly as the connected excitation level increases, to reach insignificance beyond CCSDTQ(5)$_\Lambda$. In tandem, it is also seen that basis set convergence becomes ever more rapid, with even unpolarized cc-pVDZ(p, s) yielding surprisingly small errors beyond CCSDT($Q$), and ultimately dwindling down into numerical noise.

Consequently, post-CCSD($T$) corrections tend to be evaluated in very small basis sets, and subsequently applied additively. For instance, in W4 theory [15], the $T_3 - (T)$ correction is extrapolated from cc-pVDZ and cc-pVTZ basis sets (commonly indicated by the shorthand cc-pV{D, T}Z), the ($Q$) term is evaluated in a cc-pVTZ basis set, and $T_4 - (Q)$ in just a cc-pVDZ basis set. (In W4lite theory, just CCSDT($Q$)/cc-pVDZ is done for the quadruples.)

This then leads us to the main research question of the present paper: are such corrections materially affected by BSSE corrections? The issue was raised by a reviewer of Reference [23], where we showed that post-CCSD($T$) contributions for cohesive energies of water clusters approach 1 kcal/mol for isomers of $(H_2O)_{20}$.

## 2 | Computational Details

The CCSDT($Q$) [32], CCSDT [33], CCSD($T$)$_\Lambda$ [34–37], and CCSD($T$) calculations reported in this work were carried out using a combination of the MOLPRO 2024.1 [38], CFOUR [39], and MRCC [31] electronic structure program systems, run on the CHEMFARM cluster of the Faculty of Chemistry at Weizmann. Owing to issues with inconsistent UHF solutions in the presence of ghost atoms, many of the small-molecule counterpoise data were generated using the MRCC interfaces of MOLPRO or CFOUR. For the noncovalent interactions, only closed-shell species are involved, and hence these calculations were carried out using standalone CFOUR (as memory permitted) or MRCC. The latter code was likewise used for

**TABLE 1** | Basis set convergence of RMS post-CCSD(*T*) contributions to the TAEs (kcal/mol) of the 65-molecule W4.3 subset of the W4-17 thermochemical benchmark [27].

|               | $T_3 - (T)$ | $(Q)$ | $T_4 - (Q)$ | $(Q)_\Lambda - (Q)$ | $T_4 - (Q)_\Lambda$ | $(5)_\Lambda$ | $T_5 - (5)_\Lambda$ | $(6)_\Lambda$ | $T_6 - (6)_\Lambda$ | $T_7$ |
|---|---|---|---|---|---|---|---|---|---|---|
| nihil        | 0.84 | 1.29 | 0.26  | 0.17  | 0.11  | 0.08 | 0.015 | 0.009 | 0.001 | 0.001 |
| cc-pVDZ(p, s) | 0.53 | 0.41 | 0.09  | 0.07  | 0.06  | 0.02 | 0.003 | 0.002 | 0.001 | REF   |
| cc-pVDZ(d, s) | 0.54 | 0.25 | 0.03  | 0.03  | 0.01  | 0.01 | REF   | REF   | REF   |       |
| cc-pVTZ(f, p) | 0.18 | 0.11 | 0.012 | 0.006 | 0.007 | REF  | —     | —     |       |       |
| cc-pVQZ(g, d) | 0.09 | 0.04 | REF   | REF   | REF   | —    | —     |       |       |       |
| cc-pV5Z(h, f) | 0.05 | 0.02 |       |       |       |      |       |       |       |       |
| cc-pV{Q, 5}Z | REF  | REF  |       |       |       |      |       |       |       |       |

*Note:* The underlying data were extracted from the ESI of Reference [28], except for the connected sextuples with the cc-pVDZ(d, s) basis set and septuples with the cc-pVDZ(p, s) basis set, which were calculated for the present paper using the general coupled cluster implementation [29, 30] in MRCC [31]; raw energies can be found in the present paper's ESI.

some of the additional data in Table 1, using the algorithms presented in References [29, 30].

Reference geometries for the W4-11 thermochemical benchmark [40] (which is a subset of the larger and more recent W4-17 database [27]) were taken from the ESI of the W4-17 paper and used "as is." Reference geometries for the A24 [21] and S66 [41] datasets were downloaded from the BEGDB database [42] of noncovalent interaction geometries. Using a Python program written by one of us (ES), geometries for A24x8 were generated by compressing or stretching the intermonomer distances by the eight factors {0.9, 0.95, 1.0, 1.05, 1.1, 1.25, 1.50, 2.0} from the familiar S66x8 database [41]. They are provided in the Supporting Information.

For dissociation energies, $D_e$, of diatomic molecules, we applied the standard Boys–Bernardi counterpoise definition. For the $TAE_e$ values (total atomization energies) of small polyatomics, we used the SSFC (site-site function counterpoise) generalization of Wells and Wilson [8].

The basis sets used are of the correlation consistent [43] family, ranging from cc-pV*n*Z (*n* = D, T, Q, 5) [44, 45] to aug-cc-pV*n*Z (*n* = D, T, Q, 5) [46]. The shorthand cc-pVDZ(d, s) refers to at most d and s functions, respectively, on nonhydrogen and hydrogen atoms (the full cc-pVDZ basis set would correspond to cc-pVDZ(d, p)).

## 3 | Results and Discussion

### 3.1 | Total Atomization Energies

#### 3.1.1 | Initial Check for Diatomic Molecules

RMS (root mean square) BSSE corrections for a sample of 24 heavy-atom diatomics and 10 diatomic hydrides are given in Table 2.

First of all, unsurprisingly, the effect of BSSE on the difference between CCSDT(*Q*)$_\Lambda$ and CCSDT(*Q*) is less than 0.001 kcal/mol, and can be entirely neglected. For the difference between CCSD(*T*)$_\Lambda$ and CCSD(*T*) we find 0.008 kcal/mol for the cc-pVDZ and haVDZ basis sets, which tapers down to 0.001 kcal/mol for haVQZ and cc-pV5Z.

For connected quadruples (*Q*), the RMS BSSE is less than 0.01 kcal/mol RMS even with the cc-pVDZ basis set, and smaller still for cc-pVQZ (0.003) and haVQZ (0.001 kcal/mol). We can hence conclude that BSSE on connected quadruples is negligible even for the purposes of high-accuracy work.

The situation for higher-order connected triples $T_3 - (T)$, however, is somewhat different. For the cc-pVDZ basis set, we find 0.043 kcal/mol (i.e., 0.18 kJ/mol), which however drops to 0.026 kcal/mol when diffuse functions are added, and to 0.013 kcal/mol when we move things one notch up to cc-pVTZ. Both W4 and HEAT apply cc-pV{D, T}Z extrapolations to the higher-order triples. If we do so here (with extrapolation parameters taken from table V in Reference [47]), the RMS BSSE is just 0.008 kcal/mol, which may be justifiable to neglect in view of other, larger sources of uncertainty such as residual basis set incompleteness in the CCSD(*T*) component [17]. In fact, for the cc-pV{T, Q}Z basis sets used in W4.3 theory, the BSSE will be even more negligible.

Upon comparing RMS BSSE corrections for (*T*) and for all of $T_3$ (i.e., the difference between CCSDT and CCSD), we note that for smaller basis sets like cc-pVDZ, cc-pVTZ, and haVDZ, there is more BSSE on $T_3$ than on (*T*). For larger basis sets, however, the roles are reversed.

In addition, if one considers the whole CCSDT(*Q*) – CCSD(*T*) difference, one finds partial mutual cancelation for BSSE for the larger basis sets, since the differential BSSE effects on $T_3 - (Q)$ and (*Q*) pull in opposite directions.

The bottom line for thermochemical applications appears to be that BSSE contributions are negligible for even high-accuracy work. Does this still bear out for polyatomics, or for noncovalent interactions?

#### 3.1.2 | Small Polyatomics

While we would not be able to carry out cc-pV5Z, let alone haV5Z CCSDT(*Q*) calculations on polyatomics, Table 3 presents results with smaller basis sets for a subset of about three dozen triatomics from the W4-11 thermochemical benchmark [40]. Naturally, everything becomes larger in absolute

**TABLE 2** | RMS BSSE corrections (kcal/mol) to post-CCSD(T) $D_e$ contributions for a set of 24 AB and 10 AH diatomics.

|            | (T)   | (T)$_\Lambda$ − (T) | $T_3$ − (T) | $T_3$ | (Q)   | T(Q) − (T) | (Q)$_\Lambda$ − (Q) |
|------------|-------|---------------------|-------------|-------|-------|------------|---------------------|
| cc-pVDZ    | 0.286 | 0.008               | 0.043       | 0.329 | 0.004 | 0.045      | 0.000               |
| cc-pVTZ    | 0.170 | 0.005               | 0.013       | 0.183 | 0.006 | 0.019      | 0.000               |
| cc-pV{D, T}Z[a] |   |                     | 0.008       |       | 0.008 |            |                     |
| cc-pVQZ    | 0.071 | 0.002               | 0.006       | 0.067 | 0.003 | 0.004      | 0.000               |
| cc-pV5Z    | 0.034 | 0.001               | 0.006       | 0.028 | 0.002 | 0.004      | 0.000               |
| haVDZ      | 0.214 | 0.008               | 0.026       | 0.238 | 0.002 | 0.026      | 0.000               |
| haVTZ      | 0.077 | 0.003               | 0.008       | 0.070 | 0.004 | 0.006      | 0.000               |
| haVQZ      | 0.037 | 0.001               | 0.010       | 0.028 | 0.001 | 0.009      | 0.000               |

[a]Extrapolation exponents from table 5 of Reference [47].

**TABLE 3** | RMS BSSE corrections (kcal/mol) to post-CCSD(T) TAE$_e$ contributions for a set of triatomics.

|            | (T)   | (T)$_\Lambda$ − (T) | $T_3$ − (T) | $T_3$ | (Q)   | T(Q) − (T) | (Q)$_\Lambda$ − (Q) |
|------------|-------|---------------------|-------------|-------|-------|------------|---------------------|
| cc-pVDZ    | 0.612 | 0.018               | 0.093       | 0.704 | 0.008 | 0.098      | 0.001               |
| cc-pVTZ    | 0.353 | 0.024               | 0.039       | 0.389 | 0.015 | 0.053      | 0.003               |
| w/o ClOO   | 0.353 | 0.020               | 0.036       | 0.388 | 0.015 | 0.050      | 0.002               |
| cc-pV{D, T}Z[a] |  |                     | 0.017       |       | 0.019 |            |                     |
| cc-pVQZ    | 0.143 | 0.028               | 0.017       | 0.139 | 0.006 |            | 0.004               |
| haVDZ      | 0.443 | 0.052               | 0.052       | 0.487 | 0.005 |            | 0.006               |
| haVTZ      | 0.171 | 0.024               | 0.017       | 0.168 | 0.009 |            | 0.001               |

[a]Extrapolation exponents from table 5 of Reference [47]. ClOO excluded.

numbers. Nevertheless, the same basic tendencies are seen as for the diatomics:

- BSSE on (Q) is basically insignificant and on (Q)$_\Lambda$ − (Q) wholly so.
- BSSE on $T_3$ − (T) skirts the 0.1 kcal/mol edge for cc-pVDZ, but tapers down to 0.04 for cc-pVTZ, and becomes negligible with the cc-pV{D, T}Z extrapolation practiced in W4 theory and HEAT.

We hence conclude that these thermochemical protocols require no modification to account for post-CCSD(T) BSSE unless one targets an accuracy that is likely unattainable with W4- and HEAT-type approaches.

And once again, substituting haV$n$Z for cc-pV$n$Z cuts BSSE in half.

### 3.2 | Noncovalent Interactions

#### 3.2.1 | Small Noncovalent Dimers: The A24x8 Dataset

Counterpoise corrections data for the A24x8 dataset are summarized in Table 4.

Noncovalent interactions are very different in their behavior from atomization energies, in that for most noncovalent complexes, in that MP2 is already a decent to good starting point (except for π-stacking and related). Thus, CCSD-MP2 and (T) are commonly evaluated using relatively small basis sets (see, e.g., References [41, 48] and references therein).

One might thus reasonably expect that post-CCSD(T) contributions will be proportionally much smaller. Admittedly, of course, the A24x8 systems are quite small, and hence post-CCSD(T) contributions might be somewhat less picayune in larger noncovalent complexes, especially at compressed geometries.

The RMS ΔBSSE values are even tinier in absolute terms: 0.002 kcal/mol for $T_3$ − (T) and 0.003–0.004 for (Q). For smaller basis sets, these are still nontrivial fractions of the actual Δ$D_e$ contributions. Therefore, any post-CCSD(T) corrections obtained with *very* small basis sets, such as the unpolarized double zeta cc-pVDZ(no d), need to be regarded with some caution.

For a more reasonable cc-pVTZ basis set, ΔBSSE represents about 8% of the ΔCP [$T_3$ − (T)] and 13% of Δ(Q).

#### 3.2.2 | Not-So-Small Noncovalent Complexes: The S66 Dataset

The aforementioned analysis is open to the criticism that the systems in A24 are quite small and not necessarily representative of what one might see in a real-life application. In contrast, the well-known S66 benchmark [41] consists of dimers of biomolecular building blocks interacting in different ways (hydrogen bonding, π-stacking, pure London dispersion, and mixed-influence). As such, it contains larger systems such as benzene dimer (both parallel-displaced and T-shaped, systems **24** and **47**, respectively), uracil dimer (both Watson-Crick **17** and π-stacked **26**), pentane and neopentane dimers (systems **34** and **36**, respectively).

**TABLE 4** | RMS BSSE corrections (kcal/mol) to post-CCSD(T) $D_e$ contributions for the A24x8 set of noncovalent interactions.

|  | RMS ΔCP | | | RMS ΔCP/RMS Δ$D_e$ | | |
| --- | --- | --- | --- | --- | --- | --- |
|  | (T) | $T_3 - (T)$ | (Q) | (T) | $T_3 - (T)$ | (Q) |
| cc-pVDZ(no d) | 0.060 | 0.003 | 0.004 | 0.890 | 0.205 | 0.239 |
| cc-pVDZ | 0.067 | 0.002 | 0.003 | 0.759 | 0.162 | 0.208 |
| cc-pVTZ | 0.049 | 0.001 | 0.003 | 0.290 | 0.082 | 0.134 |
| cc-pVQZ | 0.027 | 0.002 |  | 0.127 | 0.080 |  |
| haVDZ | 0.049 | 0.002 |  | 0.297 | 0.140 |  |
| haVTZ | 0.016 | 0.003 |  | 0.074 | 0.147 |  |
| haVQZ | 0.007 |  |  | 0.029 |  |  |
| haV5Z | 0.003 |  |  | 0.014 |  |  |

**TABLE 5** | RMS BSSE corrections (kcal/mol) to post-CCSD(T) $D_e$ contributions for the S66 set of larger noncovalent complexes.

|  | cc-pVDZ(p, s) | | cc-pVDZ(d, s) | |
| --- | --- | --- | --- | --- |
| Difference | RMSDiff(A, B) | $N_{systems}$ | RMSDiff(A, B) | $N_{systems}$ |
| CCSD−nihil | 3.382 | 66 | 3.290 | 66 |
| CCSD(T)−CCSD | 0.211 | 66 | 0.232 | 66 |
| CCSDT-3−CCSD(T) | 0.013 | 66 | 0.018 | 66 |
| CCSD(T)$_\Lambda$−CCSD(T) | 0.009 | 66 | 0.010 | 66 |
| CCSDT−CCSD(T) | 0.008 | 66 | 0.013 | 64[a] |
| CCSDT−CCSD(T)$_\Lambda$ | 0.007 | 66 | 0.006 | 64 |
| CCSDT−CCSDT-3 | 0.007 | 66 | 0.007 | 64 |
| CCSDT(Q)−CCSDT | 0.007 | 59 | 0.008 | 25 |
| CCSDT(Q)−CCSDT-3 | 0.011 | 59 | 0.014 | 25 |
| CCSDT(Q)−CCSD(T) | 0.003 | 59 | 0.005 | 25 |

[a]Missing S66 systems **41** uracil-pentane and **43** uracil-neopentane.

For this dataset, we will alas have to limit ourselves to the cc-pVDZ(no p on H), a.k.a., cc-pVDZ(d, s), basis set. We were able to obtain full CCSDT BSSE corrections for 64 out of 66 systems, and CCSDT(Q) for about two dozen. (It bears reiterating that, while the dimers often posed memory or computation time requirements that exceeded our available resources, the evaluation of counterpoise corrections does *not* require the dimers $AB$, only the monomers in the full dimer basis set $A(B)$ and $B(A)$, as well as naturally the monomers in their own basis set.)

Some relevant statistics can be found in Table 5. Even though with this small basis set, the BSSE correction at the CCSD(T) level is quite hefty, the *differential* BSSE correction to $T_3 − (T)$ is surprisingly modest, 0.018 kcal/mol RMS. In fact, the lion's share of even this small difference is recovered at the CCSDT-3 level [49–51]. This approximate coupled cluster approach neglects the $T_3$ term in the $T_3$ amplitude equations, thus reducing computation time scaling with system size from the $O(n_{occ.}^3 N_{virt.}^5)$ of full CCSDT to the same $O(n_{occ.}^3 N_{virt.}^4)$ as CCSD(T). (The difference between CCSDT and CCSDT-3 starts in fifth order in many-body perturbation theory [52, 53], with the leading term $E_{TT}^{[5]}$. The difference between CCSDT-3 and CCSD(T), on the other hand, has the leading term $E_{TQ}^{[5]}$ resulting from the action of the disconnected quadruples $\hat{T}_2^2/2$ on the connected triples amplitudes $T_3$.)

A still more economical approximation is offered by CCSD(T)$_\Lambda$ [34–37] which is only $O(n_{occ.}^2 N_{virt.}^4)$ in the iterations, followed by a single $O(n_{occ.}^3 N_{virt.}^4)$ step. Its cost premium over CCSD(T) is just in the need to also solve for the "left-hand eigenvectors" aside from the CCSD "right-hand" solution (which approximately doubles overall CPU time).

While we admittedly do not have as many data points for ($Q$) as for $T_3 - (T)$, for the available ones the $\Delta$CP is just 0.008 kcal/mol RMS. What is more, $\Delta\text{CP}[T_3 - (T)]$ and $\Delta\text{CP}[(Q)]$ have opposite signs (like the underlying contributions) and cancel each other to a large degree. As a result, the cumulative $\Delta\text{CP}[\text{CCSDT}(Q) - \text{CCSD}(T)]$ is just a measly 0.006 kcal/mol, which can be regarded as negligible by any reasonable standard.

If we remove all polarization functions from cc-pVDZ, we are left with just a split-valence basis set, and most of the CCSDT($Q$) calculations come within reach. To our astonishment, we found that the differential BSSEs on $T_3 - (T)$, ($Q$), and CCSDT($Q$)–CCSD($T$) remain equally tiny.

## 4 | Conclusions

In response to our research question, we can conclude the following:

1. For high-accuracy computational thermochemistry, particularly total atomization energies obtained at the W4 or HEAT levels, $T_3 - (T)$ with the cc-pVDZ basis set carries a small but noticeable BSSE.
2. This is effectively removed, however, by the extrapolation of $T_3 - (T)$ from cc-pV{D, T}Z basis sets.
3. BSSE on ($Q$) may be regarded as negligible in a thermochemistry context.
4. For noncovalent interactions, BSSE on both $T_3 - (T)$ and ($Q$) is insignificant even for basis sets as small as cc-pVDZ, and besides is subject to a degree of mutual cancelation between $T_3 - (T)$ and ($Q$).

**Author Contributions**

M.S. brought on the subject matter. V.F. performed the calculations and curated the data for the TAE BSSE corrections. E.S. carried out the A24x8 calculations and curated the relevant data. E.S. and J.M.L.M. shared the S66 calculations. J.M.L.M. oversaw the project and wrote the first draft, while all authors contributed to the completed manuscript.

**Acknowledgments**

This work was supported by the Israel Science Foundation (grant 1969/20) and by the Uriel Arnon Memorial Center for AI research into smart materials. J.M.L.M. thanks the Quantum Theory Project at the University of Florida and its head, Prof. John F. Stanton, for their hospitality. The authors would like to thank Drs. Peter R. Franke, Branko Ruscic, and Nisha Mehta as well as Profs. A. Daniel Boese (KFU Graz, Austria) and Leslie Leiserowitz (Weizmann) for inspiring discussions, and Dr. James R. Thorpe and Prof. Devin A. Matthews (Southern Methodist U., Dallas, TX) for fixing a particularly vexing memory allocation bug in CFOUR, thus removing an insurmountable obstacle for the S66 calculations. All calculations were carried out on the ChemFarm HPC cluster of the Weizmann Institute Faculty of Chemistry.

**Conflicts of Interest**

The authors declare no conflicts of interest.

**Data Availability Statement**

The data that support the findings of this study are available in the Supporting Information of this article.

**References**

1. S. F. Boys and F. Bernardi, "The Calculation of Small Molecular Interactions by the Differences of Separate Total Energies. Some Procedures With Reduced Errors," *Molecular Physics* 19, no. 4 (1970): 553–566.

2. B. Brauer, M. K. Kesharwani, and J. M. L. Martin, "Some Observations on Counterpoise Corrections for Explicitly Correlated Calculations on Noncovalent Interactions," *Journal of Chemical Theory and Computation* 10, no. 9 (2014): 3791–3799.

3. L. A. Burns, M. S. Marshall, and C. D. Sherrill, "Comparing Counterpoise-Corrected, Uncorrected, and Averaged Binding Energies for Benchmarking Noncovalent Interactions," *Journal of Chemical Theory and Computation* 10, no. 1 (2014): 49–57.

4. C. Hättig, W. Klopper, A. Köhn, and D. P. Tew, "Explicitly Correlated Electrons in Molecules," *Chemical Reviews* 112, no. 1 (2012): 4–74.

5. L. Kong, F. A. Bischoff, and E. F. Valeev, "Explicitly Correlated R12/F12 Methods for Electronic Structure," *Chemical Reviews* 112, no. 1 (2012): 75–107.

6. S. Ten-no and J. Noga, "Explicitly Correlated Electronic Structure Theory From R12/F12 Ansätze," *Wiley Interdisciplinary Reviews: Computational Molecular Science* 2, no. 1 (2012): 114–125.

7. A. Halkier, H. Koch, P. Jørgensen, O. Christiansen, I. M. Beck Nielsen, and T. Helgaker, "A Systematic *Ab Initio* Study of the Water Dimer in Hierarchies of Basis Sets and Correlation Models," *Theoretical Chemistry Accounts* 97, no. 1–4 (1997): 150–157.

8. B. H. Wells and S. Wilson, "van der Waals Interaction Potentials: Many-Body Basis Set Superposition Effects," *Chemical Physics Letters* 101, no. 4–5 (1983): 429–434.

9. N. J. DeYonker, T. R. Cundari, and A. K. Wilson, "The Correlation Consistent Composite Approach (ccCA): An Alternative to the Gaussian-*n* Methods," *Journal of Chemical Physics* 124, no. 11 (2006): 114101.

10. C. Peterson, D. A. Penchoff, and A. K. Wilson, "Prediction of Thermochemical Properties Across the Periodic Table: A Review of the Correlation Consistent Composite Approach (ccCA) Strategies and Applications," in *Annual Reports in Computational Chemistry*, vol. 12, ed. D. A. Dixon (Amsterdam, Netherlands: Elsevier, 2016), 3–45.

11. A. Tajti, P. G. Szalay, A. G. Császár, et al., "HEAT: High Accuracy Extrapolated *Ab Initio* Thermochemistry," *Journal of Chemical Physics* 121, no. 23 (2004): 11599–11613.

12. Y. J. Bomble, J. Vázquez, M. Kállay, et al., "High-Accuracy Extrapolated *Ab Initio* Thermochemistry. II. Minor Improvements to the Protocol and a Vital Simplification," *Journal of Chemical Physics* 125, no. 6 (2006): 064108.

13. M. E. Harding, J. Vázquez, B. Ruscic, A. K. Wilson, J. Gauss, and J. F. Stanton, "High-Accuracy Extrapolated *Ab Initio* Thermochemistry. III. Additional Improvements and Overview," *Journal of Chemical Physics* 128, no. 11 (2008): 114111.

14. J. H. Thorpe, C. A. Lopez, T. L. Nguyen, et al., "High-Accuracy Extrapolated *Ab Initio* Thermochemistry. IV. A Modified Recipe for Computational Efficiency," *Journal of Chemical Physics* 150, no. 22 (2019): 224102.

15. A. Karton, E. Rabinovich, J. M. L. Martin, and B. Ruscic, "W4 Theory for Computational Thermochemistry: In Pursuit of Confident Sub-kJ/mol Predictions," *Journal of Chemical Physics* 125, no. 14 (2006): 144108.

16. A. Karton, P. R. Taylor, and J. M. L. Martin, "Basis Set Convergence of Post-CCSD Contributions to Molecular Atomization Energies," *Journal of Chemical Physics* 127, no. 6 (2007): 064104.

17. N. Sylvetsky, K. A. Peterson, A. Karton, and J. M. L. Martin, "Toward a W4-F12 Approach: Can Explicitly Correlated and Orbital-Based *Ab Initio* CCSD(T) Limits Be Reconciled?," *Journal of Chemical Physics* 144, no. 21 (2016): 214101.

18. D. A. Dixon, D. Feller, and K. A. Peterson, "Chapter One — A Practical Guide to Reliable First Principles Computational Thermochemistry Predictions Across the Periodic Table," in *Annual Reports in Computational Chemistry*, vol. 8, ed. R. A. Wheeler (Amsterdam, Netherlands: Elsevier, 2012), 1–28.

19. D. Feller, K. A. Peterson, and B. Ruscic, "Improved Accuracy Benchmarks of Small Molecules Using Correlation Consistent Basis Sets," *Theoretical Chemistry Accounts* 133 (2014): 1407.

20. V. Fishman, E. Semidalas, and J. M. L. Martin, "Basis Set Extrapolation from the Vanishing Counterpoise Correction Condition," *Journal of Physical Chemistry A* 128, no. 35 (2024): 7462–7470.

21. J. Řezáč and P. Hobza, "Describing Noncovalent Interactions Beyond the Common Approximations: How Accurate Is the "Gold Standard," CCSD(T) at the Complete Basis Set Limit?," *Journal of Chemical Theory and Computation* 9, no. 5 (2013): 2151–2155.

22. T. Schäfer, A. Irmler, A. Gallo, and A. Grüneis, "Understanding Discrepancies of Wavefunction Theories for Large Molecules," 2024, https://arxiv.org/abs/2407.01442.

23. G. Santra, M. Shepelenko, E. Semidalas, and J. M. L. Martin, "Is Valence CCSD(T) Enough for the Binding of Water Clusters? The Isomers of $(H_2O)_6$ and $(H_2O)_{20}$ as a Case Study," 2023, https://arxiv.org/abs/2308.06120.

24. K. Raghavachari, G. W. Trucks, J. A. Pople, and M. Head-Gordon, "A Fifth-Order Perturbation Comparison of Electron Correlation Theories," *Chemical Physics Letters* 157, no. 6 (1989): 479–483.

25. J. D. Watts, J. Gauss, and R. J. Bartlett, "Coupled-Cluster Methods With Noniterative Triple Excitations for Restricted Open-Shell Hartree–Fock and Other General Single Determinant Reference Functions. Energies and Analytical Gradients," *Journal of Chemical Physics* 98, no. 11 (1993): 8718–8733.

26. A. D. Boese, M. Oren, O. Atasoylu, J. M. L. Martin, M. Kállay, and J. Gauss, "W3 Theory: Robust Computational Thermochemistry in the kJ/mol Accuracy Range," *Journal of Chemical Physics* 120, no. 9 (2004): 4129–4141.

27. A. Karton, N. Sylvetsky, and J. M. L. Martin, "W4-17: A Diverse and High-Confidence Dataset of Atomization Energies for Benchmarking High-Level Electronic Structure Methods," *Journal of Computational Chemistry* 38, no. 24 (2017): 2063–2075.

28. E. Semidalas, A. Karton, and J. M. L. Martin, "W4Λ: Leveraging Λ Coupled-Cluster for Accurate Computational Thermochemistry Approaches," *Journal of Physical Chemistry. A* 128, no. 9 (2024): 1715–1724.

29. M. Kállay and P. R. Surján, "Higher Excitations in Coupled-Cluster Theory," *Journal of Chemical Physics* 115, no. 7 (2001): 2945–2954.

30. M. Kállay and J. Gauss, "Approximate Treatment of Higher Excitations in Coupled-Cluster Theory," *Journal of Chemical Physics* 123, no. 21 (2005): 214105.

31. M. Kállay, P. R. Nagy, D. Mester, et al., "The MRCC Program System: Accurate Quantum Chemistry From Water to Proteins," *Journal of Chemical Physics* 152, no. 7 (2020): 074107.

32. Y. J. Bomble, J. F. Stanton, M. Kállay, and J. Gauss, "Coupled-Cluster Methods Including Noniterative Corrections for Quadruple Excitations," *Journal of Chemical Physics* 123, no. 5 (2005): 054101.

33. J. Noga and R. J. Bartlett, "The Full CCSDT Model for Molecular Electronic Structure," *Journal of Chemical Physics* 86, no. 12 (1987): 7041–7050.

34. J. F. Stanton and J. Gauss, "A Simple Correction to Final State Energies of Doublet Radicals Described by Equation-of-Motion Coupled Cluster Theory in the Singles and Doubles Approximation," *Theoretica Chimica Acta* 93, no. 5 (1996): 303–313.

35. T. D. Crawford and J. F. Stanton, "Investigation of an Asymmetric Triple-Excitation Correction for Coupled-Cluster Energies," *International Journal of Quantum Chemistry* 70, no. 4–5 (1998): 601–611.

36. S. A. Kucharski and R. J. Bartlett, "Noniterative Energy Corrections Through Fifth-Order to the Coupled Cluster Singles and Doubles Method," *Journal of Chemical Physics* 108, no. 13 (1998): 5243–5254.

37. S. A. Kucharski and R. J. Bartlett, "Sixth-Order Energy Corrections With Converged Coupled Cluster Singles and Doubles Amplitudes," *Journal of Chemical Physics* 108, no. 13 (1998): 5255–5264.

38. H.-J. Werner, P. J. Knowles, F. R. Manby, et al., "The Molpro Quantum Chemistry Package," *Journal of Chemical Physics* 152, no. 14 (2020): 144107.

39. D. A. Matthews, L. Cheng, M. E. Harding, et al., "Coupled-Cluster Techniques for Computational Chemistry: The CFOUR Program Package," *Journal of Chemical Physics* 152, no. 21 (2020): 214108.

40. A. Karton, S. Daon, and J. M. L. Martin, "W4-11: A High-Confidence Benchmark Dataset for Computational Thermochemistry Derived From First-Principles W4 Data," *Chemical Physics Letters* 510, no. 4–5 (2011): 165–178.

41. J. Řezáč, K. E. Riley, and P. Hobza, "S66: A Well-Balanced Database of Benchmark Interaction Energies Relevant to Biomolecular Structures," *Journal of Chemical Theory and Computation* 7, no. 8 (2011): 2427–2438.

42. J. Řezáč, P. Jurečka, K. E. Riley, et al., "Quantum Chemical Benchmark Energy and Geometry Database for Molecular Clusters and Complex Molecular Systems (www.begdb.com): A Users Manual and Examples," *Collection of Czechoslovak Chemical Communications* 73, no. 10 (2008): 1261–1270.

43. T. H. Dunning, Jr., K. A. Peterson, and D. E. Woon, "Basis Sets: Correlation Consistent Sets," in *Encyclopedia of Computational Chemistry*, ed. P. von Ragué Schleyer (New York: Wiley & Sons, 1998).

44. T. H. Dunning, Jr., "Gaussian Basis Sets for Use in Correlated Molecular Calculations. I. The Atoms Boron Through Neon and Hydrogen," *Journal of Chemical Physics* 90, no. 2 (1989): 1007–1023.

45. D. E. Woon and T. H. Dunning, Jr., "Gaussian Basis Sets for Use in Correlated Molecular Calculations. III. The Atoms Aluminum Through Argon," *Journal of Chemical Physics* 98, no. 2 (1993): 1358–1371.

46. R. A. Kendall, T. H. Dunning, Jr., and R. J. Harrison, "Electron Affinities of the first-Row Atoms Revisited. Systematic Basis Sets and Wave Functions," *Journal of Chemical Physics* 96, no. 9 (1992): 6796–6806.

47. A. Karton, "Effective Basis Set Extrapolations for CCSDT, CCSDT(Q), and CCSDTQ Correlation Energies," *Journal of Chemical Physics* 153, no. 2 (2020): 024102.

48. B. Brauer, M. K. Kesharwani, S. Kozuch, and J. M. L. Martin, "The S66x8 Benchmark for Noncovalent Interactions Revisited: Explicitly Correlated *Ab Initio* Methods and Density Functional Theory," *Physical Chemistry Chemical Physics* 18, no. 31 (2016): 20905–20925.

49. J. D. Watts and R. J. Bartlett, "The Coupled-Cluster Single, Double, and Triple Excitation Model for Open-Shell Single Reference Functions," *Journal of Chemical Physics* 93, no. 8 (1990): 6104–6105.

50. M. Urban, J. Noga, S. J. Cole, and R. J. Bartlett, "Towards a Full CCSDT Model for Electron Correlation," *Journal of Chemical Physics* 83, no. 8 (1985): 4041–4046.

51. J. Noga, R. J. Bartlett, and M. Urban, "Towards a Full CCSDT Model for Electron Correlation. CCSDT-$n$ Models," *Chemical Physics Letters* 134, no. 2 (1987): 126–132.

52. S. A. Kucharski and R. J. Bartlett, "Fifth-Order Many-Body Perturbation Theory and Its Relationship to Various Coupled-Cluster Approaches," *Advances in Quantum Chemistry* 18 (1986): 281–344.

53. Y. He, Z. He, and D. Cremer, "Comparison of CCSDT-$n$ Methods With Coupled-Cluster Theory With Single and Double Excitations and Coupled-Cluster Theory With Single, Double, and Triple Excitations in Terms of Many-Body Perturbation Theory—What is the Most Effective Triple-Excitation Method?," *Theoretical Chemistry Accounts* 105, no. 3 (2001): 182–196.

**Supporting Information**

Additional supporting information can be found online in the Supporting Information section.